\documentclass[aps,reprint,showpacs]{revtex4-1}
\usepackage{graphicx}
\usepackage{mathtools} \usepackage{amsmath}
\usepackage{epstopdf}
\usepackage{dcolumn} \usepackage{graphicx} \usepackage{bm} \usepackage[colorlinks]{hyperref}
\hypersetup{
	citecolor=blue,
	linkcolor=blue,
	urlcolor=blue}

\begin{document}
	
\title{Structural, magnetic, electrical and calorimetric properties of hexagonal Gd$_5$Sb$_3$}

\title{Magnetic phase separation and strong AFM nature of hexagonal Gd$_5$Sb$_3$}

\author{S. Shanmukharao Samatham} \email{sssrao@phy.iitb.ac.in} 
\author{Akhilesh Kumar Patel}
\author{K. G. Suresh}
\affiliation{Magnetic Materials Laboratory, Department of Physics, Indian Institute of Technology Bombay, Mumbai 400076, Maharashtra, India}

\date{\today}

\begin{abstract}

We report on the combined results of structural, magnetic, transport and calorimetric properties of Mn$_5$Si$_3$-type hexagonal Gd$_5$Sb$_3$. With decreasing temperature, it exhibits a ferromagnetic-like transition at 265 K, N\'{e}el transition at 95.5 K and a spin-orientation transition at 62 K. The system is found to be in AFM state down to 2 K in a field of 70 kOe. Magnetic phase coexistence is not noticeable despite large positive Curie-Weiss temperature. Instead low-temperature AFM and high-temperature FM-like phases are separated in large temperature. Temperature-magnetic field ($H$-$T$) phase diagram reveals field-driven complex magnetic phases. Within the AFM phase, the system is observed to undergo field-driven spin-orientation transitions. Field-induced tricritical and quantum critical points appear to be absent due to strong AFM nature and by the intervention of FM-like state between PM and AFM states. Electrical resistivity along with large Sommerfeld parameter suggests metallic nature.

\end{abstract}

\maketitle

\section{Introduction}

Gd-based compounds have been given due concern in the quest of large magnetocaloric effect (MCE). Various doping combinations have been tried ever since the discovery of large room temperature MCE in Gd. Later, R$_5$Ge$_4$ (R = Gd and M = Ge, Si) compounds have been recognized as promising MCE candidates with giant MCE (Gd$_5$Ge$_2$Si$_2$ \cite{PecharskyPhysRevLett.78.4494,GschneidnerRepProgPhys2005}), along with rich physical phenomena such as frozen magnetostructural first-order magnetic phase transition \cite{ChandraPhysRevB.79.052402}. In recent times, belonging to the similar category but structurally different compounds R$_5$M$_3$ (M=Ge and Sb), are actively investigated for their interesting novel ground states \cite{KUSHWAHA20121824,MajiClusterGlassJPCM2011,MajiMCEJAP2012,MajiMHJumpsEPL2010,NirmalaPrGe3NdGe3JAP2011,TSUTAOKAPhysicaBNd5Ge32010,TamizhavelPhysRevBPrGe3TbGe32009,NagaiGd5Sb3JMMM2007,SvitlykJSSC2008}. Most of these compounds are found to crystallize in Mn$_5$Si$_3$-type hexagonal structure with P63/mcm (No. 193) space group \cite{BuschowPSSB1967}. The unit cell consists of five layers of atoms in c-axis direction. In this, R atom is located in two non-equivalent positions 4d (1/3, 2/3, 0) and 6g ($x_\mathrm{R}$, 0, 0.25) whereas M atom is situated in 6g ($x_\mathrm{M}$, 0, 0.25) position in which $x_\mathrm{R}$ and $x_\mathrm{M}$ are element specific. The low-temperature properties are reported to be affected mainly by the disorder caused by the presence of two non-equivalent sites of R atom; 6g and 4d. The interaction between the two layers (4d and 6g positions of R atom) plays an important role in determining the magnetic properties of these compounds.

Rare-earth antimonides R$_5$Sb$_3$ (R = Nd, Gd, Ho and Sm) have been reported to order antiferromagnetically with varying N\'{e}el temperatures ($T_\mathrm{N}$), with an exception to La$_5$Sb$_3$ and Y$_5$Sb$_3$ \cite{YakinthosJMMM1983}. Gd$_5$Sb$_3$ is reported to show paramagnetic (PM) to antiferromagnetic (AFM) transition around 110 K. Abdusaljamove et al. reported complex magnetic structure of Gd$_5$Sb$_3$ \cite{AbdusaljamovaJLCM1986} with Curie temperature $T_\mathrm{C} \sim$ 260 K, Curie-Weiss temperature $\theta_\mathrm{CW} \sim$ 263 K, effective magnetic moment $\mu_\mathrm{eff} \sim$ 8.46/Gd$^{+3}$ and saturation moment $M_\mathrm{s} \sim$ 32.5 $\mu_\mathrm{B}/\mathrm{f.u.}$ (under 4.2 K and 200 kOe). Nagai et al. have reported three transitions at $T_\mathrm{C} \sim$ 187 K, $T_\mathrm{N} \sim$ 100 K and a ferrimagnetic transition below 50 K \cite{NagaiGd5Sb3JMMM2007} with $\theta_\mathrm{CW} \sim$ 266.8 K and $\mu_\mathrm{eff} \sim$ 10.9 $\mu_\mathrm{B}$/Gd$^{+3}$.

Manifestation of properties of the compounds that possess single magnetic phase transition such as PM-AFM or PM-FM, is somewhat predictable under the effect of external fields $H$. In either case, where $T_\mathrm{N}$ or $T_\mathrm{C}$ is suppressed by $H$ down to $T$ = 0, a quantum critical point is achieved. In a case where applied field induces the ferromagnetism, it leads to either metamagnetic (sudden) or spin-flop (gradual) transition. However, the scenario in compounds with multiple magnetic transitions is non-trivial in the sense that coexistence of phases leads to first order phase transition, zero-field ground state with weak AFM correlations leads to field-induced tricritical point etc.

The current study presents the detailed investigation of Gd$_5$Sb$_3$ using the combined results of structure, magnetization, specific heat and electrical resistivity. It shows temperature-driven multiple transitions from PM-FM ($T_\mathrm{C}$), FM to AFM ($T_\mathrm{C}$) and a low temperature spin-orientation like transition ($T_\mathrm{t}$). The effect of magnetic field on these phases is discussed. Eventually, temperature-field ($T$-$H$) phase diagram, depicting field-induced magnetic changes in different magnetic phases, is presented.

\section{Experimental Methods}

Polycrystalline Gd$_5$Sb$_3$ is prepared by arc melting the constituent elements (of purity at least 99.99\%), taken in the stoichiometric ratio, under the inert Ar gas atmosphere. The ingot was remelted by flipping several times to ensure homogeneity. The weight loss during melting is less than 1\%. The sample is annealed at 1273 K for seven days. X-ray diffraction pattern at room temperature is collected using PANalytical X'Pert PRO X-ray diffractometer on a powder specimen using Cu-K$_\mathrm{\alpha}$ radiation. The sample is found to be phase pure with no detectable traces of secondary phase, crystallizing in Mn$_5$Si$_3$-type hexagonal crystal structure. Rietveld refinement of the same is shown in Fig. \ref{fig:XRD1}. Magnetization is measured using commercial Superconducting Quantum Interference Device-Vibrating Sample Magnetometer (SQUID-VSM). The measurements are carried out in isothermal and iso-field modes. Resistivity (by dc-linear four probe method) and specific heat (by relaxation calorimetric method) are carried out using 9 T/ 2 K - Physical Property Measurement System (Quantum Design).

\section{Results}

\begin{figure}
	\centering
	\includegraphics[width=0.5\linewidth, height=0.5\linewidth]{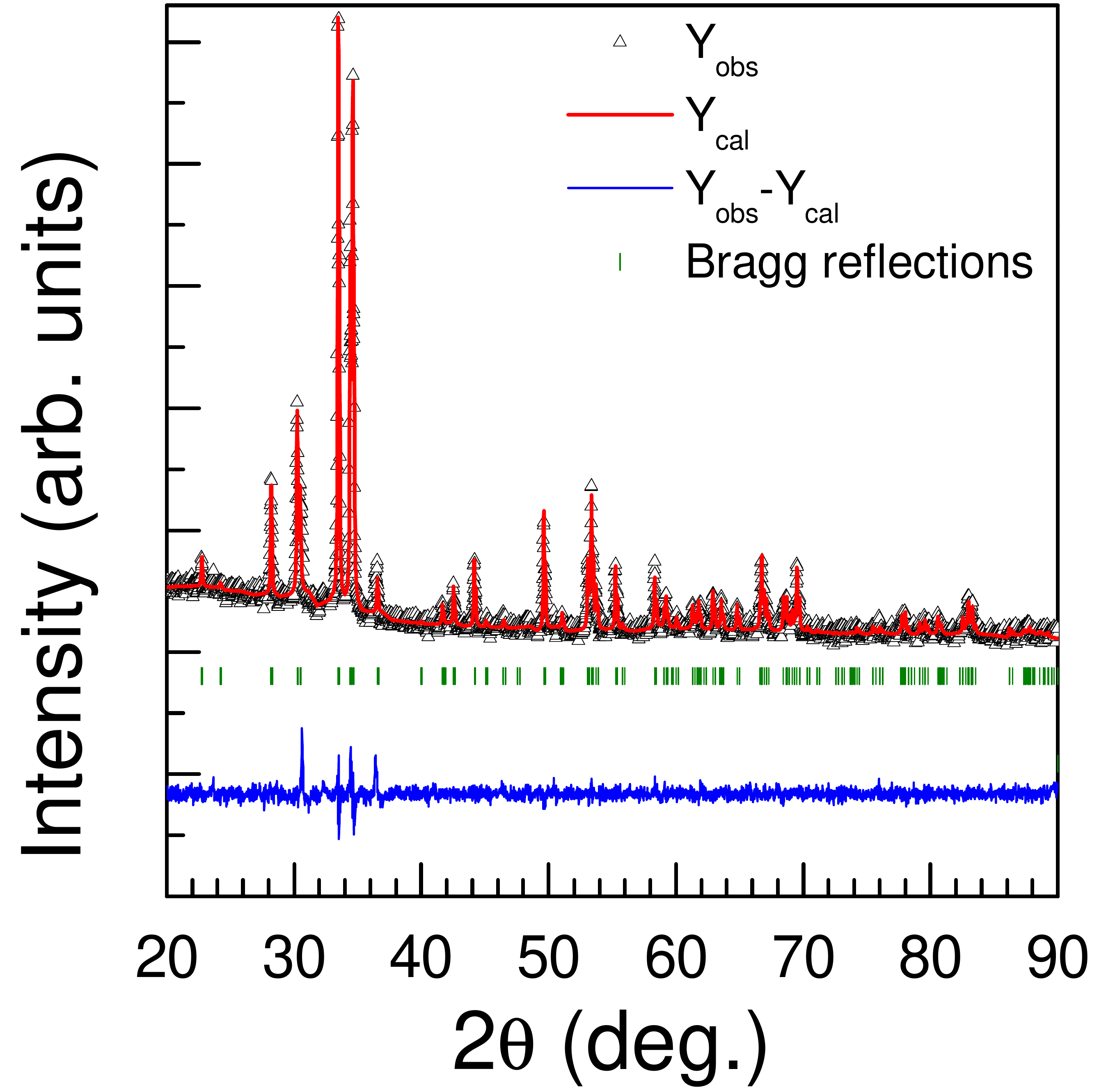}
	\caption{Rietveld refined room temperature X-ray diffraction pattern of Gd$_5$Sb$_3$. Experimental data shows good agreement with the generated pattern using P 63/m c m space group. The estimated lattice parameters are $a (=b) \sim$ 9.02 \AA~ and $c=$ 6.32 \AA.}
	\label{fig:XRD1}
\end{figure}

\begin{table}
	\caption{Wyckoff positions of Gd$_5$Sb$_3$ after refinement}
	\centering
	\begin{tabular}{c c c c c}
		\hline
		Position & Atom& \textit{x} & \textit{y} & \textit{z} \\
		\hline
		4d & Gd & 0.3333 & 0.6667 & 0 \\
		\hline
		6g & Gd & 0.24852 & 0 & 0.25 \\
		\hline
		6g & Sb & 0.61253 & 0 & 0.25\\
		\hline
	\end{tabular}
	\label{table:Wyckoff}
\end{table}

\begin{figure}
	\includegraphics[width=\linewidth, height=0.5\linewidth]{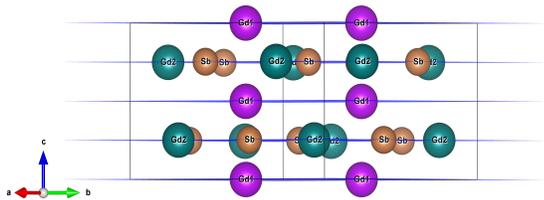}
	\caption{The unit cell of Gd$_5$Sb$_3$ consisting of five layers along \textit{c}-axis direction. Top, middle and bottoms layers are with 4d (Gd$_1$) atoms and rest of the layers are occupied by 6g (Gd$_2$ and Sb) atoms.}
	\label{fig:c-direction}
\end{figure}

\begin{figure}
	\includegraphics[width=\linewidth, height=0.5\linewidth]{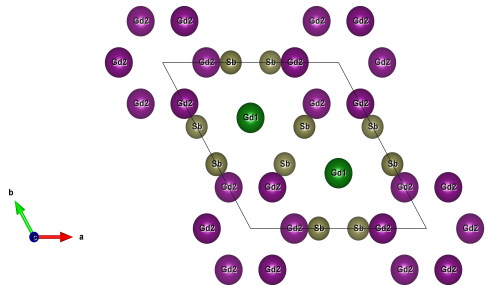}
	\caption{ab-plane projection of unit cell with extended along corners for Gd$_5$Sb$_3$.}
	\label{fig:ab-projection}
\end{figure}

Shown in Fig. \ref{fig:XRD1}(a) is the room temperature X-ray diffraction pattern along with the Rietveld refinement (using FullProf suit \cite{FullProfSuitePhysicaB1993}) using space group P6/3mcm (No: 193). The experimental intensities (Y$_\mathrm{obs}$) are in excellent agreement with the generated intensities (Y$_\mathrm{cal}$). The lattice parameters are $a (= b) \sim$ 9.02 \AA~ and $c$ = 6.32 \AA~which are in good agreement with the previous reports \cite{RiegerActaR5Sb31968,AbdusaljamovaJLCM1986}. Table \ref{table:Wyckoff} shows the refined parameters along with atomic positions. Shown in Fig. \ref{fig:c-direction} is the Gd$_5$Sb$_3$ unit cell. It comprises of five layers of atoms along the \textit{c}-axis. Bottom, middle and top layers are occupied by 4d (Gd$_1$) positions and rest of the layers are filled with 6g (Gd$_2$ and Sb) positions. Fig. \ref{fig:ab-projection} shows the ab-plane projected view of Gd$_5$Sb$_3$ structure. The atoms are labeled as per the respective sites; Gd$_1$ at 4d, Gd$_2$ at 6g and Sb at 6g positions. At each corner, Gd$_2$ atoms form two opposite triangles (resembling hexagonal from top-view) with a side of 3.881(6) \AA. The nearest neighbor Gd$_1$-Gd$_1$ (4d-4d) inter-atomic distance is 3.16 \AA~and inter-layer Gd$_2$-Gd$_1$ (6g-4d) inter-atomic distance is 3.80 \AA.

\begin{figure}
	\centering
	\includegraphics[width=\linewidth, height=0.5\linewidth]{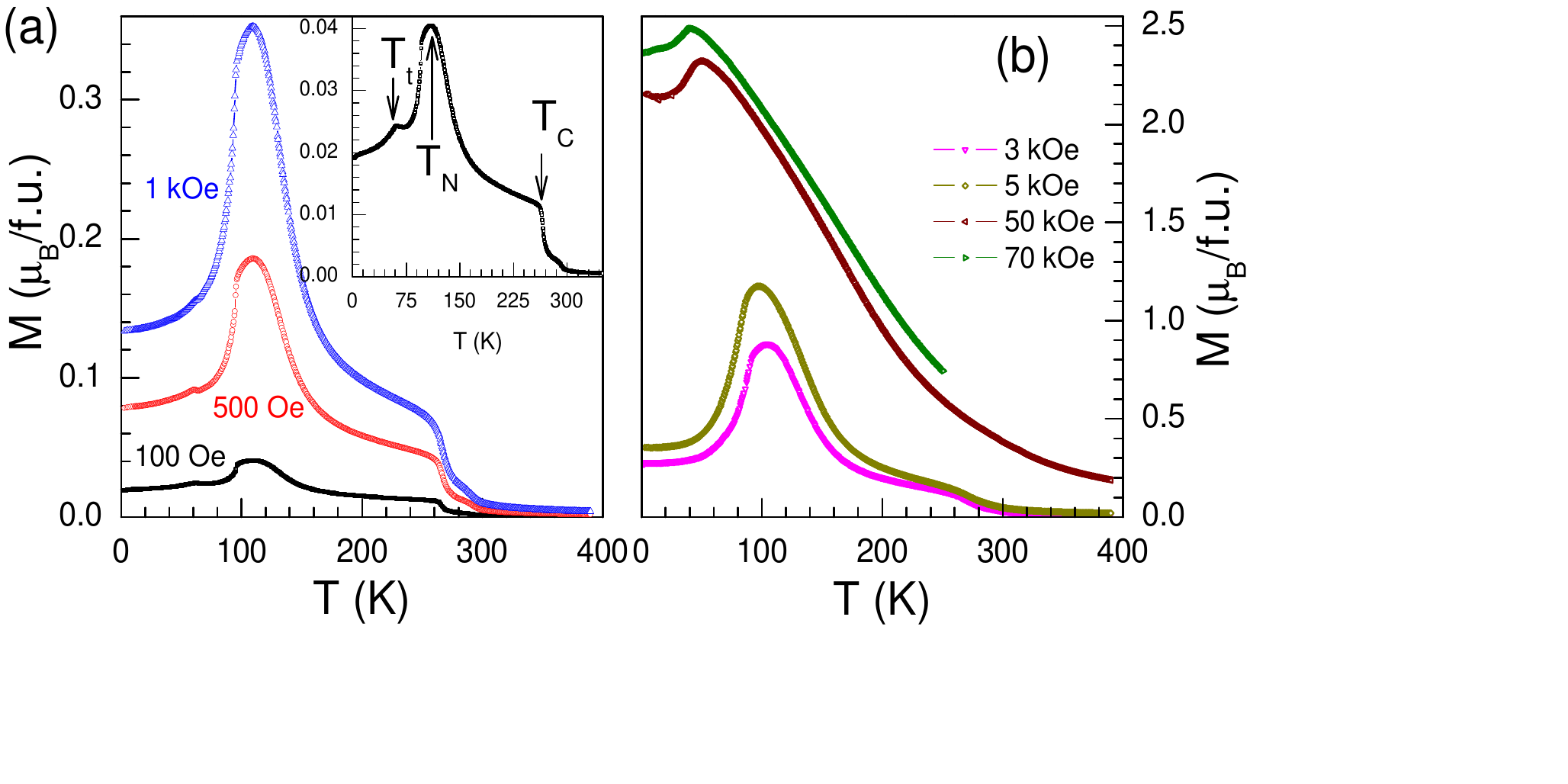}
	\caption{Temperature dependence of magnetization of Gd$_5$Sb$_3$ in few representative magnetic fields. (a) $M(T)$ in 0.1, 0.5 and 1 kOe. The magnetic transitions are indicated by arrows (inset). $T_\mathrm{N}$ is observed to shift to low-$T$ with $H$ while $T_\mathrm{C}$ shifts to high-$T$.}
	\label{fig:MT}
\end{figure}

\begin{figure*}
	\centering
	\includegraphics[width=0.7\linewidth, height=0.233\linewidth]{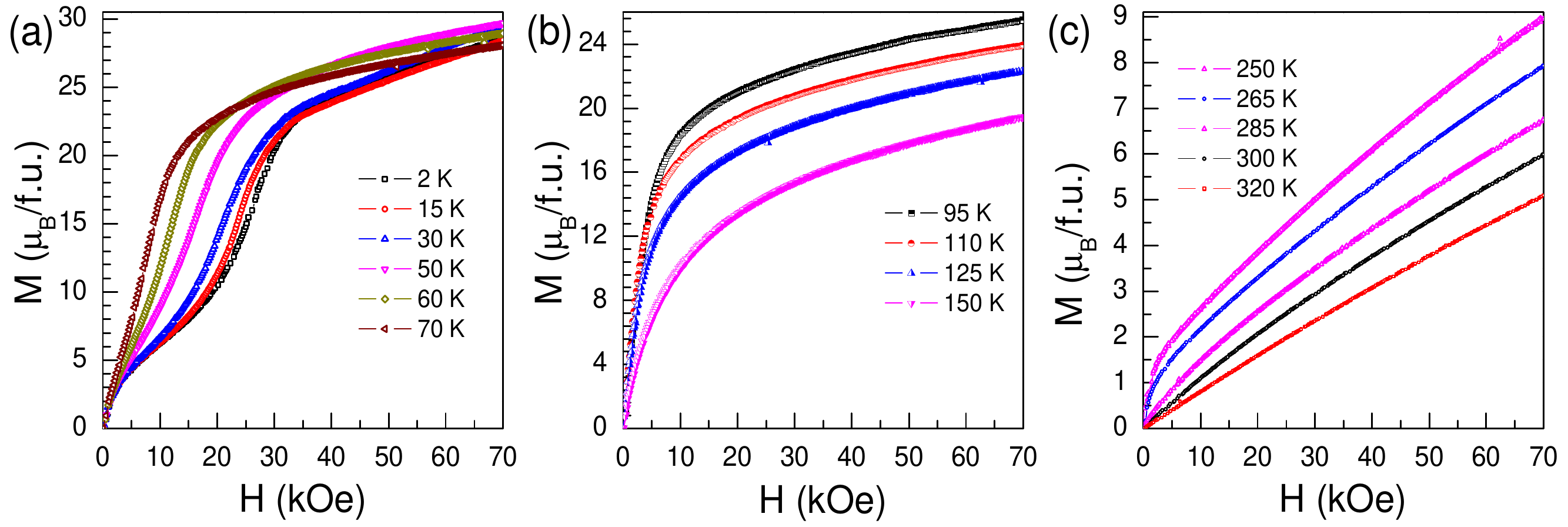}
	\caption{Magnetization isotherms of Gd$_5$Sb$_3$ from 0 to 70 kOe, measured at labeled temperatures. (a) Below $T-\mathrm{t}$ field-induced transition is noticed while (b) above $T_\mathrm{N}$, ferromagnetic-like smooth increase of $M(H)$ is discernible. (c) At high temperatures i.e., in the paramagnetic region $M$ goes linearly with $H$.}
	\label{fig:MH1Q}
\end{figure*}

Figure \ref{fig:MT} shows the temperature dependence of magnetization from 390 to 2 K, measured in zero-field cooled (ZFC) protocol, in few selected magnetic fields. Inset of Fig. \ref{fig:MT}(a) shows $M(T)$ in 100 Oe. Magnetization increases steeply below 400 K with a slope change at $T_\mathrm{C} \sim$ 265 K resembling ferromagnetic transition. After attaining a maximum around 100 K, magnetization falls rapidly below 95.5 K discerning an antiferromagnetic transition at $T_\mathrm{N}$. In addition, a small peak around 62 K (= $T_\mathrm{t}$) is noticed. It is observed that $T_\mathrm{N}$-peak significantly increases and shifts to low-temperatures with increasing $H$ while $T_\mathrm{C}$ shifts to high-$T$. $T_\mathrm{t}$ shifts to low-$T$ by simultaneously smearing out. Shown in Fig. \ref{fig:MH1Q} are the magnetization isotherms $M(H)$ at few representative temperatures. A cursory look at the isotherms reveals that magnetization exhibits distinct field-driven phase crossovers at low temperatures up to 95 K. Between 95 K and 285 K, magnetization increases smoothly with $H$, similar to FM behavior. Almost linear increase of $M$ with $H$ is discernible above 285 K.

\section{Discussion}

\begin{figure}
\includegraphics[width=0.6\linewidth, height=0.6\linewidth]{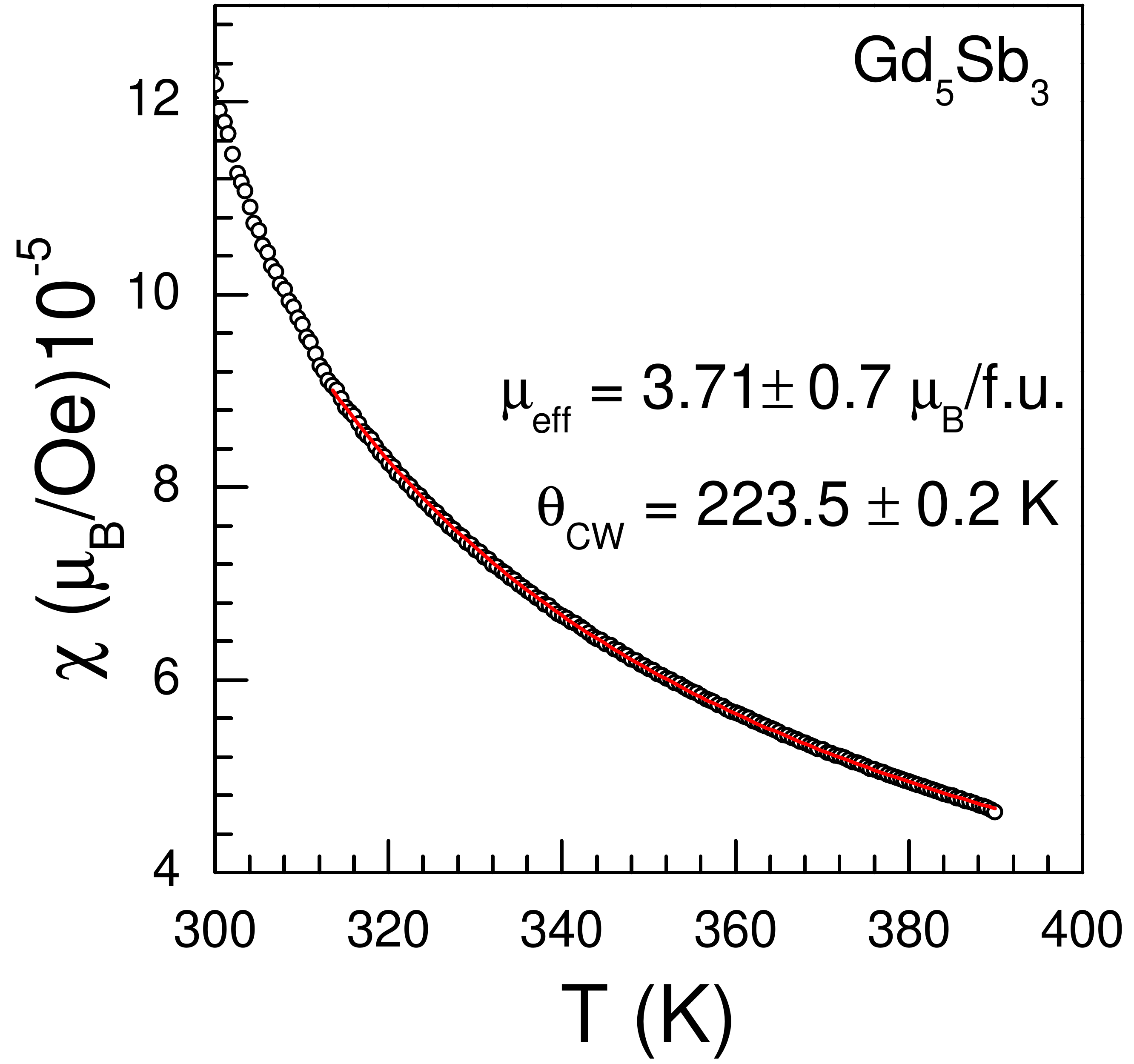}
\caption{Temperature dependence of dc-susceptibility $\chi_\mathrm{dc}(T)$ in 100 Oe. A fit of Curie-Weiss law Eq. \ref{Eq:CWLaw} in the paramagnetic region yields a $\theta_\mathrm{CW}$ of about 224 K and effective moment of about 3.7 $\mu_\mathrm{eff}$/f.u.}
\label{fig:Chi}
\end{figure}

Figure \ref{fig:Chi} shows the dc-susceptibility ($\chi_\mathrm{dc} = M/H$) in 100 Oe from 390 to 310 K i.e., in the paramagnetic state. Curie-Weiss fit is used to estimate the effective moment ($\mu_\mathrm{eff}$) and Curie-Weiss temperature $\theta_\mathrm{CW}$.

\begin{equation}
\chi(T) = \dfrac{N_\mathrm{A}\mu^2_\mathrm{eff}}{3k\mathrm{_B}(T-\theta_\mathrm{CW})}
\label{Eq:CWLaw}
\end{equation}
where $N_\mathrm{A}$ and $k_\mathrm{B}$ are the Avogadro's and Boltzmann constants, $\theta_\mathrm{CW}$ is Curie-Weiss temperature. Thus estimated $\theta_\mathrm{CW}$ and $\mu_\mathrm{eff}$ are 223.5 $\pm$ 0.2 K and 3.71 $\pm$ 0.7 $\mu_\mathrm{B}$/Gd$^{+3}$, respectively. Curie-Weiss temperature is large and positive which is in agreement with the literature \cite{AbdusaljamovaJLCM1986}. However, the observed effective moment is smaller than that of Gd$^{+3}$ free ion (7.94 $\mu_\mathrm{B}$ with $J =$ 7/2) and previously reported values 8.46 $\mu_\mathrm{B}$/Gd$^{+3}$ \cite{AbdusaljamovaJLCM1986} and 10.9 $\mu_\mathrm{B}$/Gd$^{+3}$ \cite{NagaiGd5Sb3JMMM2007}. While large positive Curie-Weiss temperature indicates the ferromagnetic correlations, the low effective moment may indicate the non-contribution of some of the Gd moments by screening effects. Suppression of the moment due to crystal field is also ruled out as it is Gd. In addition, $M(T)$ measurements are also performed in field-cooled cooling (FCC) and field-cooled warming (FCW) modes. No considerable hysteresis between FCC and FCW curves is observed, ruling out the first order behavior between high-$T$ FM-like to low-$T$ AFM phases.

\begin{figure*}
	\centering
	\includegraphics[width=\linewidth, height=0.5\linewidth]{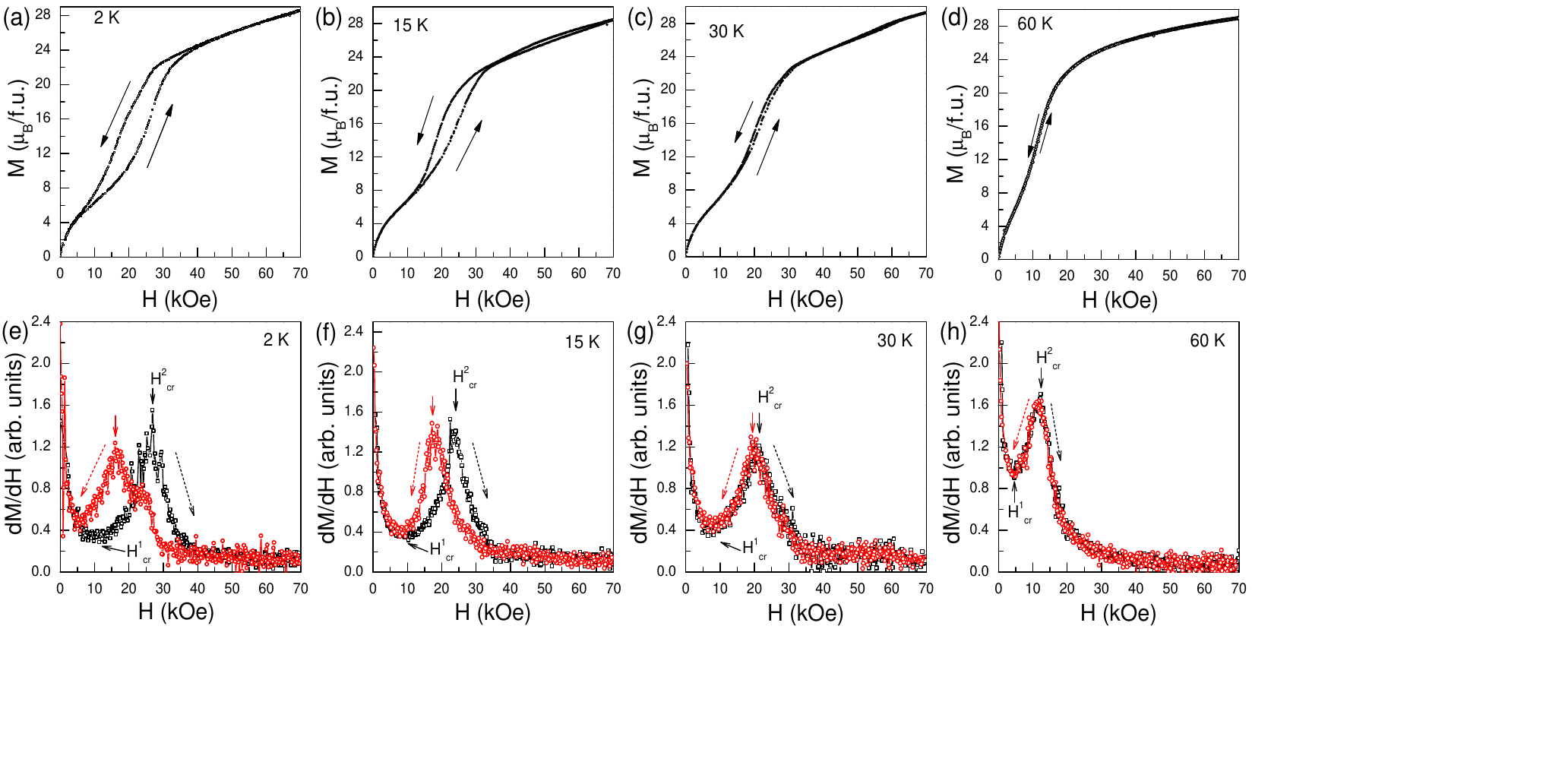}
	\caption{(a-d) Magnetization isotherms of Gd$_5$Sb$_3$ measured in increasing and decreasing magnetic fields at few representative temperatures and (e-h) their corresponding $dM/dH$ plots. Below spin orientation temperature ($T_\mathrm{t} \sim$ 62 K), the compound shows field-induced hysteresis loop, shown in $dM/dH$ as two distinct peaks in red hallow circles and black hallow squares.}
	\label{fig:Hysteresis}
\end{figure*}

In order to understand the magnetic states in Gd$_5$Sb$_3$, it  is worth knowing various types of phenomena resulting from the competing magnetic interactions. The resultant depends on not only the strength of interactions but also the nature of high- and low-$T$ magnetic phases. In most of the rare-earth based compounds, generally one comes across two types of magnetic phases \textit{viz}., AFM and FM and a competition between them. For instance, we discuss the types of magnetic phenomena in compounds with dual transitions and no phase coexistence. First let us consider the case with a transition from high-$T$ AFM to low-$T$ FM phase. With increasing $H$, FM transition is shifted to high-$T$ while that of AFM phase is shifted to low-$T$. Apparently, with increasing magnetic field strength, two simultaneous processes occur; stabilization of low-$T$ FM at high temperatures and suppression of AFM phase to low temperatures (an otherwise extension of PM phase to low temperatures). As a result, the zero-field thermal separation $\Delta T$ (= $\left|T_\mathrm{N}-T_\mathrm{C}\right|$) between the phases is reduced in finite fields. In other words, thermal ($k_\mathrm{B}\Delta T$) and magnetic ($\mu H$) energies act cooperatively with each other and consequently establish FM state by ceasing the high-$T$ AFM phase in relatively low-fields. The temperature and field at which PM-AFM transition ceases to exist allowing a continuous PM-FM transition is generally noted as field-induced tricritical point.

\begin{figure}
	\includegraphics[width=\linewidth, height=0.5\linewidth]{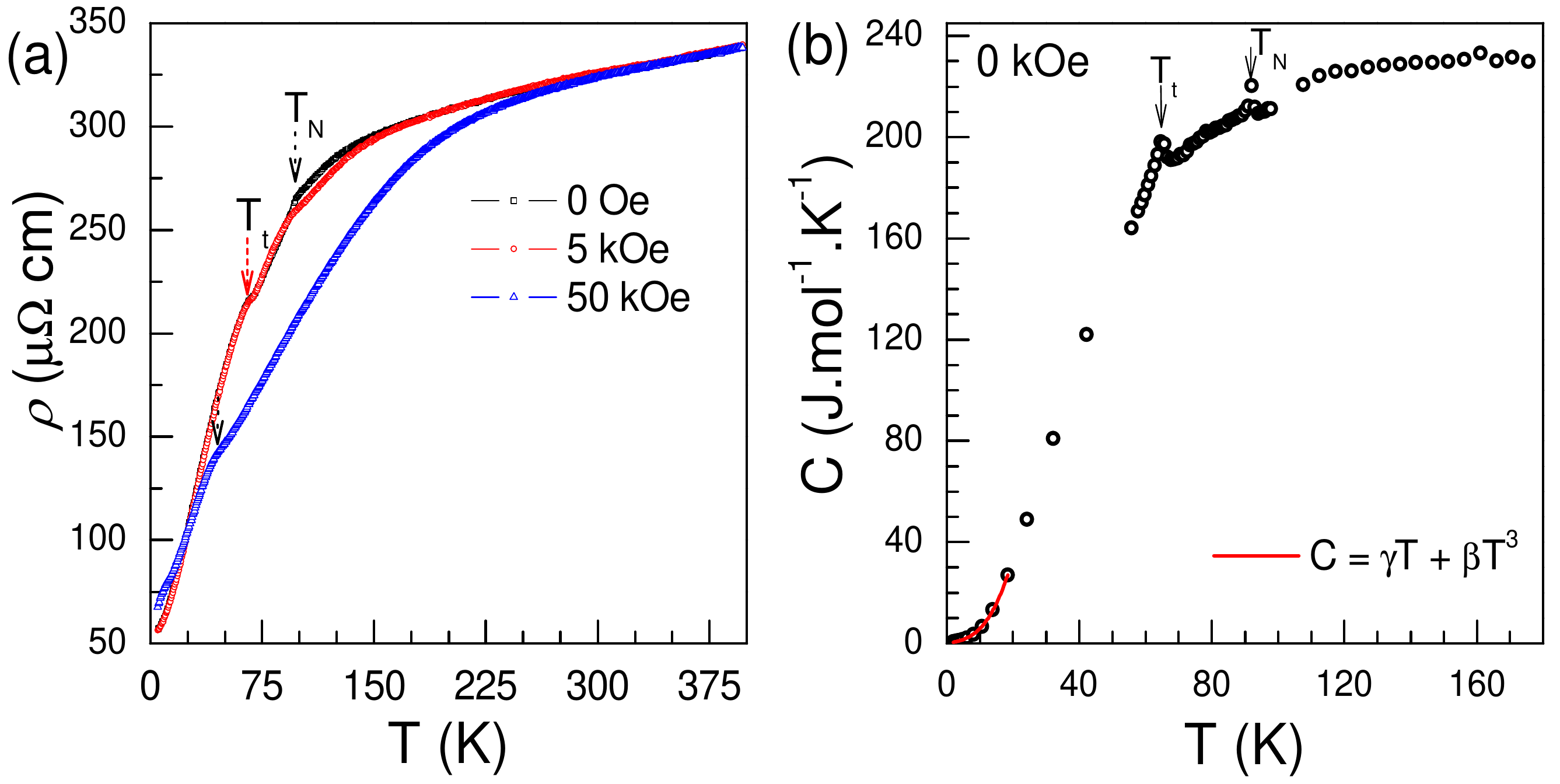}
	\caption{Temperature variation of (a) resistivity and (b) specific heat. The dual transitions $T_\mathrm{N}$ and $T_\mathrm{t}$ are indicated by arrows. Specific below 20 K is fitted to the equation $C = \gamma T + \beta T^2$ and the estimated Sommerfeld parameter is about 215 mJ.mol$^{-1}$.K$^{-1}$.} 
	\label{fig:ResSpheat}
\end{figure}

The scenario is just the opposite in the case of a transition from high-$T$ FM and low-$T$ AFM phase. With increasing field strength, $T_\mathrm{C}$ and $T_\mathrm{N}$ continue to increase in opposite temperature directions. Low-$T$ AFM phase is suppressed to further low temperatures and high-$T$ FM phase is expanded to low temperatures at the cost of AFM phase. Zero-field $\Delta T$ increases with field. As a result, $k_\mathrm{B}\Delta T$ and $\mu H$ oppose each other, requiring high magnetic fields to stabilize FM phase down to the lowest temperature. Nevertheless, the compounds (for example Gd$_3$Co \cite{SamathamGd3CoJMMM2017} and PrCuSi \cite{StrydomPrCuSiEJP2010}) undergoing a transition from PM to AFM with prevailing FM correlations in the paramagnetic limit (identifiable by positive $\theta_\mathrm{CW}$) are found to be unstable against applied $H$. The weakly correlated AFM phase is susceptible to $H$ and eventually converts to FM phase in presence of relatively low fields. However, irrespective of the nature of high- and low-$T$ magnetic phases, in the context of phase-coexistence (AFM and FM) phenomena the properties are governed by the competing interactions of AFM and FM magnetic clusters formed in zero-field. Applied $H$ tries to break the antiferromagnetically interacting clusters, leading to FM phase growth. Such a scenario may either lead to first-order kinetic arrest or second-order spin glass states \cite{RoyMGRSGPRB2009}.

Now, we attempt to understand the magnetic scenario of Gd$_5$Sb$_3$. As seen in the Figs. \ref{fig:MT}(a\&b), Gd$_5$Sb$_3$ belongs to the latter case with high-$T$ FM-like and low-$T$ AFM. The separation of phases in temperature is noticeable from the increase of $T_\mathrm{C}$ and decrease of $T_\mathrm{N}$. To know more about the magnetic nature of the compound, magnetization isotherms are measured in five quadrants ($0 \rightarrow H_\mathrm{max} \rightarrow 0 \rightarrow -H_\mathrm{max} \rightarrow 0 \rightarrow H_\mathrm{max}$). For clarity, only first quadrant is shown in Fig. \ref{fig:MH1Q}, for magnetizing and demagnetizing fields since the positive and negative quadrants are found to be symmetric. At 2 K, two field-driven transitions are observed at $H_\mathrm{cr}^1 \sim$ 11.4 kOe and $H_\mathrm{cr}^2 \sim$ 26.7 kOe. $H_\mathrm{cr}^2$ is attributed to AFM to high moment FM-like spin-flop transition whereas $H_\mathrm{cr}^1$ is a spin-reorientation transition. No saturation is observed up to 70 kOe. Variations of $H^1_\mathrm{cr}$ and $H^2_\mathrm{cr}$ with temperature are represented by plotting $dM/dH$ vs. $H$, Figs. \ref{fig:Hysteresis}(e-f). The hysteresis between $H^1_\mathrm{cr}$ and $H^2_\mathrm{cr}$ is perceptible from the low $H^2_\mathrm{cr}$ in decreasing fields (peak at 15.98 kOe represented by red hallow circles). The non-saturation of $M(H)$ is because of an incomplete transformation of AFM phase at 2 K, as shown in $M(T)$ the $\lambda$-type N\'{e}el transition at 37.6 K. With increasing temperature (i.e., moving towards $T_\mathrm{N}$), $H_\mathrm{cr}^1$ and $H_\mathrm{cr}^2$ decrease monotonically, suggesting the ceasing of field-driven phase transition and hysteresis simultaneously at $T_\mathrm{N}$. Further, above $T_\mathrm{N}$, a smooth increase of $M$ with $H$ is observed, which is in consonance with the ferromagnetic-like behavior between $T_N \le T \le T_\mathrm{C}$.

Further to understand the transport and calorimetric behavior, electrical resistivity and specific heat of Gd$_5$Sb$_3$ are measured. Shown in Fig. \ref{fig:ResSpheat}(a) is the temperature dependence of resistivity $\rho(T)$ from 2-375 K in a few representative fields, 5 and 50 kOe. The system is metallic down to 2 K with a residual resistivity $\rho_0$ of about 48 $\mu\Omega$cm. Relatively large residual resistivity ratio ($\rho_\mathrm{300 K}/\rho_{0} \simeq$ 7) implies the good quality of the prepared polycrystalline sample. Zero-field $\rho(T)$ shows kinks at $T_\mathrm{N}$ and $T_\mathrm{t}$. In agreement with the $M(T)$ measurements, magnetic field suppression of the transitions is observed in field-resistivity curves. Figure \ref{fig:ResSpheat}(b) depicts specific heat $C(T)$ from 2-285 K, showing $T_\mathrm{N}$ and $T_\mathrm{t}$ respectively at 92 K and 65 K. $C(T)$ is fitted using the equation $C = \gamma T + \beta T^3$ for $T \le$ 20 K. The estimated Sommerfeld coefficient $\gamma \sim$ 215 mJ.mol$^{-1}$.K$^{-1}$, implying the metallic character. Debye temperature $\theta_\mathrm{D} \sim$ 161 K.

\begin{figure}
	\includegraphics[width=0.6\linewidth, height=0.6\linewidth]{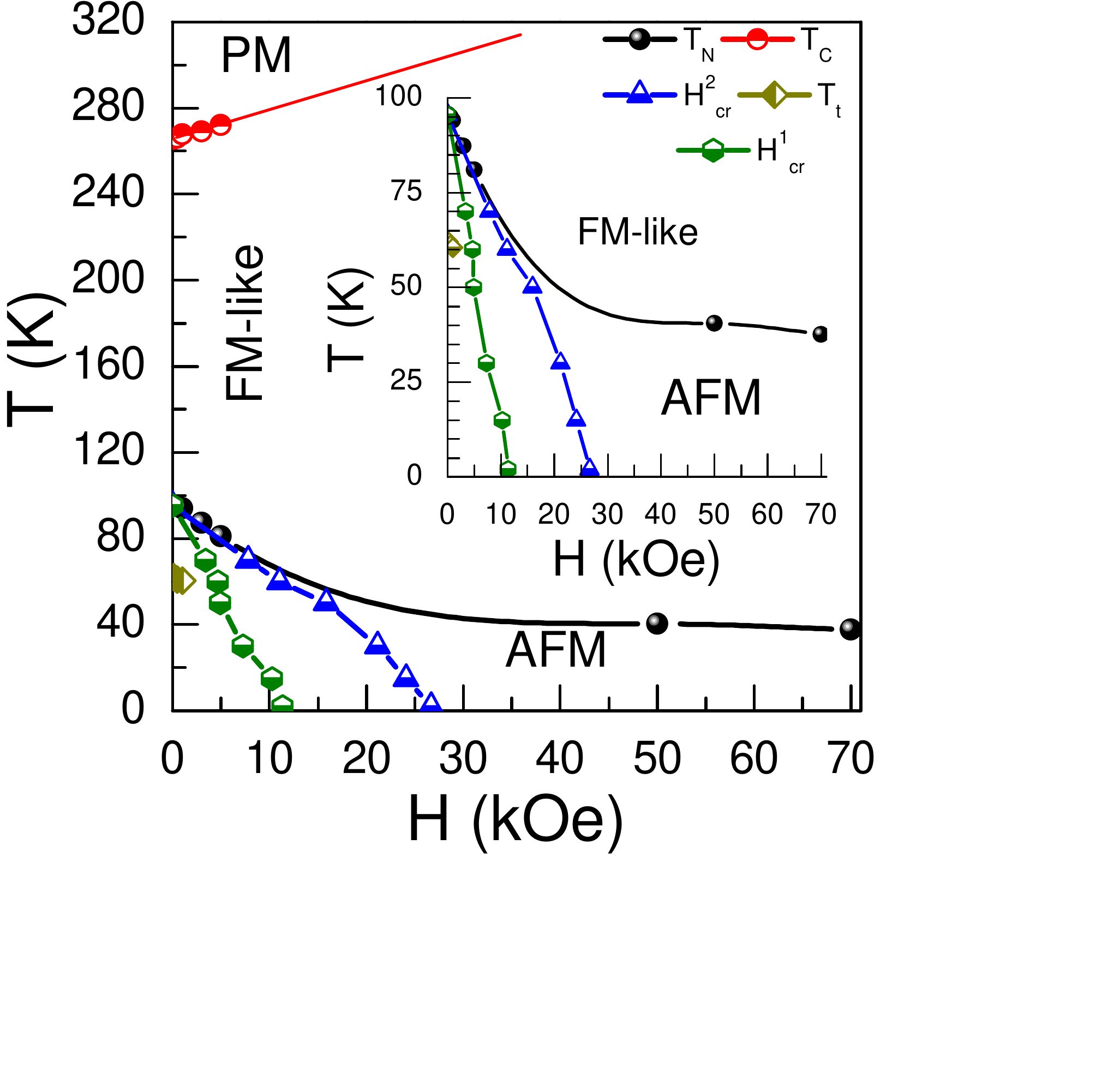}
	\caption{Temperature-field ($T$-$H$) phase diagram of Gd$_5$Sb$_3$, showing paramagnetic (PM), ferromagnetic (FM)-like and AFM phases with decreasing temperature. $T_\mathrm{C}$ increases with field whereas $T_\mathrm{N}$ decreases. Solid red is guide to an eye. Inset: field-induced dual transitions in AFM region using $H^1_\mathrm{cr}$ and $H^2_\mathrm{cr}$ boundaries. $T_\mathrm{t}$ boundary ends at 3 kOe.} 
	\label{fig:PD}
\end{figure}

Temperature-magnetic field ($T$-$H$) phase diagram is shown in Fig. \ref{fig:PD}, which is constructed on the basis of field-variation of $T_\mathrm{C}$ and $T_\mathrm{N}$ along with temperature-variation of critical fields $H_\mathrm{cr}^1$ and $H_\mathrm{cr}^2$. FM and AFM regions are separated in temperature. With increasing fields, FM expands to low-$T$ by suppressing AFM phase. The system is observed to undergo two spin-reorientation transitions at 2 K within 30 kOe. However, it is noticed that higher fields are required to completely suppress the AFM state down to 2 K. The system avoids field-induced tricritical and quantum critical points are due to strong AFM characteristics and by the intervention of FM-like state between PM and AFM phases.

As mentioned earlier, no thermal hysteresis between FCC and FCW curves in the temperature range $T \in$ [$T_\mathrm{C}$, $T_\mathrm{N}$] discords the temperature-driven first order phase transition in Gd$_5$Sb$_3$. Despite large positive Curie-Weiss temperature, FM and AFM phases are found not to coexist. Instead, the magnetic phases are thermodynamically separated in zero field. A rough estimation of magnetic field needed to bring full FM nature at low temperatures is estimated by equating thermal energy $k_\mathrm{B}\Delta T$ with $\mu_\mathrm{eff} H$. For Gd$_5$Sb$_3$, with $\Delta T$ = 170 K, $H \sim$ 684 kOe is required to overcome the thermal energy barrier. However, at room temperature, the field required to saturate the magnetization is estimated to 1200 kOe. $J$ value, reverse calculated using $\mu_\mathrm{eff} = 2\sqrt{J(J+1)}$ = 3.7 $\mu_\mathrm{B}/\mathrm{Gd}^{+3}$, is close to 3/2. However, magnetic state below N\'{e}el temperature is complex in character and needs further attention. Initially, at 2 K, it undergoes field-driven dual transitions from canted to AFM and AFM to spin-flop state within external fields of 70 kOe.

\section{Conclusion}

In conclusion, we report on the combined results of structural, magnetic, transport and calorimetric properties of Gd$_5$Sb$_3$. It crystallizes in Mn$_5$Si$_3$-type hexagonal crystal structure. AFM and FM-like phase are separated in large temperature. It exhibits field-driven complex magnetic phases. No magnetic phase coexistence is observed. Large Sommerfeld coefficient and positive temperature coefficient of resistivity suggest metallic nature. Within the AFM phase, the system is observed to undergo field-driven spin-orientation transitions. Field-induced tricritical and quantum critical points appear to be absent due to the strong AFM nature and by the intervention of FM-like state between PM and AFM states. In order to reveal the complex magnetic structure of Gd$_5$Sb$_3$, magnetization measurements in high (pulsed) fields along with neutron diffraction measurements as a function of temperature are highly desirable.

\section*{Acknowledgments}

The authors acknowledge IRCC and Department of Physics for the magnetization, specific heat and resistivity facilities. SSS is supported by Indian Institute of Technology Bombay Institute Post Doctoral Program.

\bibliography{refgd5sb3}

\end{document}